\documentclass[journal,draftcls,onecolumn,12pt,twoside]{IEEEtranTCOM}
\usepackage{graphicx}
\usepackage{cite}
\usepackage{xcolor}
\normalsize
\begin{document}
%
\title{Turbulence Mitigation Scheme for Optical Communications using Orbital          Angular Momentum Multiplexing Based on Channel Coding and Wavefront Correction}
\author{Shengmei~Zhao,
        Bei~Wang,
        Li~Zhou,
        Longyan~Gong,
        Wenwen~Cheng,
        Yubo~Sheng,
        Baoyu~Zheng,~\IEEEmembership{Member,~IEEE,}

\thanks{S. Zhao, B. Wang, L. Zhou, W. Cheng, Y. Sheng and B. Zheng are with the Institute
of Signal Processing and Transmission, Nanjing University of Posts and Telecommunications,
Nanjing, China, 210003,
 e-mail: zhaosm@njupt.edu.cn.}
\thanks{L. Gong is with the Department of Applied Physics, Nanjing University of Posts
and Telecommunications, Nanjing 210003, China.
e-mail: lygong@njupt.edu.cn}}

%
%


\maketitle

\begin{abstract}
The free-space optical (FSO) communication links with orbital angular momentum (OAM) multiplexing have been demonstrated that they can largely enhance the systems' capacity without a corresponding increase in spectral bandwidth, but the performance of the system is unavoidably disturbed by atmospheric turbulence (AT). Different from the existed AT disturbance, the OAM-multiplexed systems will cause both the burst and random errors for a single OAM state carrier and the `crosstalk' interference between the different OAM states carriers. In this paper, we propose a turbulence mitigation method to improve AT tolerance of OAM-multiplexed FSO communication links.  In the proposed scheme, we use channel codes to correct the burst and random errors caused by AT for a single OAM state carrier; And we use wavefront correction method to correct the `crosstalk' interference between the different OAM states carriers. The improvements of AT tolerance are discussed by comparing the performance of OAM-multiplexed FSO communication links with or without channel coding or Shark-Hartmann wavefront correction method. The numerical simulation results show that the OAM-multiplexed FSO communication links have enhanced their AT tolerance. The usage of  channel codes and wavefront correction methods together has improved greatly the performance of OAM-multiplexed FSO communication links over atmospheric turbulence.
\end{abstract}


%
\IEEEpeerreviewmaketitle

\section{Introduction}
\IEEEPARstart{S}{ince} the beginning of the electromagnetic era, the ways to send increasingly larger amounts of information in ever-smaller periods of time is a crucial task in communication systems, especially in our big-data times \cite{TO12}. In 2012, Wang \emph{et al.} demonstrated a new high-speed free-space optical (FSO) communications scheme using orbital angular momentum (OAM) multiplexing \cite{WA12}. It was capable of reliably transferring up to $1369.6 Gbit s^{-1}$ with  spectral efficiency  $25.6 bit s^{-1}Hz^{-1}$.  This demonstrations suggest that OAM could be a useful degree of freedom for tremendously increasing the capacity of free-space communications.  
Because the theoretically unlimited values of the azimuthal mode index $\ell$ for an OAM state and the mutual orthogonality for the different OAM states \cite{WA12,DJ10}, the optical communication systems using OAM multiplexing can handle a huge amount of information and tremendously increase the capacity of communication links.
Currently, OAM multiplexing techniques have been extensively studied in free-space communication links \cite{MO07,AW10,FA11,DO11,WA11,WA11_2} and fiber \cite{BO12,BO13} communications systems.

OAM states describe the spatial distributions of wave-functions with helical phases \cite{FR08,MA06,MC11,MO07,SA08,TA12,TH07,TU96,UC10,YA11}, they are inevitably sensitive to atmospheric turbulence interference \cite{ZH13_2,PA05,AN08}. 
Both theoretical and experimental results have indicated that even a weak turbulence can disturb OAM states. Turbulence aberration is detrimental to the quality of communication links using OAM states. For a single OAM state information-carrier, there already exists several approaches to mitigate the power fading  problems caused by AT. For example, Zhao \emph{et al.} and us used Reed-Solomon (RS) codes as channel coding to improve the performance of FSO communication system and holographic ghost imaging system, independently \cite{ZH10,ZH13}. Djordjevic \emph{et al.} presented a low-density parity check (LDPC)-precoded orbital angular momentum modulation scheme over a $1km$ free-space laser communication link subject to OAM modal crosstalk induced by AT \cite{DJ10,DJ11,DJ12}. Ellerbroek \emph{et al.} and Sharma  independently showed that turbulence can be to a certain degree minimized by employing adaptive optics \cite{EL94,SH12}, and we have proposed two wavefront correction methods to mitigate the effect of AT in FSO communications \cite{ZH12}.

Different from the bit errors caused by AT for a single OAM state information-carrier,  the OAM-multiplexed FSO communication links will be disturbed by both the burst and random errors  for a single OAM state carrier and the `crosstalk' interference between the different OAM states carriers \cite{AN08}. Currently, there are a few reports on the effect of AT on OAM-multiplexed communication links and the corresponding mitigation method. In \cite{RE13}, Ren \emph{et al.} experimentally investigated the performance of OAM-multiplexed FSO communication link through emulated atmospheric turbulence. The results indicated that turbulence-induced signal fading and crosstalk could deteriorate link
performance. And in `frontiers in Optics' \cite{RE13_2}, Ren \emph{et al.} discussed a turbulence compensation scheme using a single adaptive-optics system. How to mitigate the detrimental effect of AT both from single OAM state information-carrier and multiple OAM states information-carriers  has been so far much less analyzed and reported.

In this paper, we will study a method to alleviate the damaged effect of atmospheric turbulence on the OAM-multiplexed FSO communication links using both channel coding and wavefront correction method.  We use channel codes to correct the burst and random errors on the individual OAM state information-carrier. Simultaneously, we use Shark-Hartmann wavefront correction method to mitigate the phase distortion, that produce `crosstalk' interference between the different OAM states information-carriers.
The improvements of AT tolerance of OAM-multiplexed FSO communication links using channel coding and wavefront correction method are analyzed  by comparing the performance of OAM-multiplexed FSO communication links with or without channel coding and Shark-Hartmann wavefront correction method. Meanwhile, the fluctuations caused by AT are simulated by a set of random phase screens, and spatial light modulations (SLM) are employed to control these random phase screens to the propagation beam.


 The rest of the paper is organized as follows.  In section II,  a  turbulence-mitigation scheme for a communication links using OAM-multiplexing is presented by using both channel coding and wavefront correction method. In Section III, performances of the turbulence-mitigation scheme are discussed when the OAM-multiplexed FSO communication links is with or without channel coding and Shark-Hartmann wavefront correction method. And the conclusions are given in Section IV.

\section{The turbulence mitigation scheme for OAM-multiplexed FSO communication links}

In this section, we will present the turbulence mitigation scheme for OAM-multiplexed FSO communication links, the thin sheet phase screens model for atmospheric turbulence, and two correction methods in the proposed scheme.
\subsection{The turbulence mitigation scheme}
\begin{figure}
\centering
\includegraphics{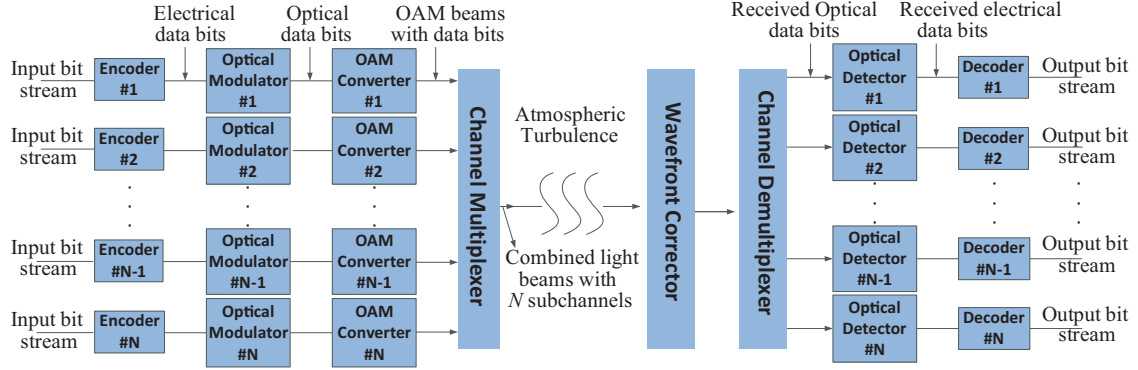}
\caption{An OAM-multiplexed  FSO communication links scheme with channel coding and wavefront correction through atmospheric turbulence.}
\label{FIG_I}
\end{figure}
The schematic diagram of the turbulence mitigation scheme for OAM-multiplexed FSO communication links is shown in Fig.~\ref{FIG_I}. At the sender, each input bit stream is first encoded to a codeword by a separate Encoder. The resulting codeword is then modulated to a Gaussian beam. The electrical data bits are transferred to the optical data bits in Optical Modulators.  Later, the Gaussian beam is converted to an OAM state beam by an OAM Converter, in which the planar phase front Gaussian beam is transferred into an information-carrying OAM beam with a helical phase front by utilizing a spatial light modulator (SLM) with a specific hologram. Finally, $N$ OAM beams are multiplexed to a superposition OAM state beam in a Channel Multiplexer and the superposition OAM state beam is transmitted through the channel with atmospheric turbulence. At the receiver, the damaged OAM superposition state beam is first corrected by Shack-Hartmann wavefront correction in Wavefront Corrector, then is de-multiplexed to obtain the different OAM states in Channel Demultiplexer. Different from the OAM multiplexing technique used in \cite{WA12}, the channel demultiplexer differentiates the different information-carrying OAM states from the superposition state beam to different lateral positions by using the method proposed in \cite{GR10}; Optical Detectors operated on each OAM state lateral positions transfer back the optical data bits to the electrical bit sequence. Finally, each bit sequence is decoded by a separate Decoder to get an output bit stream.

Depending on the parameter of a light beam, the optical modulators can be categorized into amplitude modulators, phase modulators, and polarization modulators etc.  The proposed scheme will mainly focus on the method to alleviate the damaged effect of AT on OAM multiplexing.  Hence, a simple but widely used modulation method named on-off-keyed amplitude-modulation  (OOK) is chosen in the scheme. The current of laser diode is easily controlled to produce two discrete current levels to represent the digital logical symbols. A specific spiral phase mask for producing a special OAM state has been designed in each OAM Converter for converting a Gaussian beam with planar phase front to a special information-carrying OAM beam with a helical phase front by using a SLM. Meanwhile, in the Channel De-multiplexer, the method using two static optical elements to efficiently sort OAM states of light has been used \cite{GR10}, where the optical elements perform a Cartesian to log-polar coordinate transformation, converting the helically phased light beam corresponding to OAM states into a beam with a transverse phase gradient. A subsequent Fourier transform (len) then focuses each input OAM state to a different lateral position. 
By an avalanche photodiode (APD) photo-detector at each lateral position,  the optical data bits carried by each OAM states could be transferred back to electrical data bits.
\subsection{Thin sheet phase screens model for atmospheric turbulence }
In the scheme, the turbulence aberration is simulated using thin sheet phase screens model \cite{ST04}. The spatial variation caused by AT can be approximated by several thin sheets with random phase screen $\phi(x,y)$ that modify the phase profile of the propagating beam. The initial beam is propagated using the standard Fourier optics techniques to move between $x$-space and $k$-space.  The amplitude of the electric field  $U_{1-}(x,y)$ just prior to the first phase screen would be,
\begin{equation}
U_{1-}(x,y)=FFT^{-1}[FFT(U_0(x,y))\widehat{U}(k_x,k_y)],
\label{EQ_I}
\end{equation}
where $U_0(x,y)$ is the amplitude of the electric field at the beginning, $\widehat{U}(k_x,k_y)$ is the propagation operator in $k$-space, $FFT$ and $FFT^{-1}$ represent fast Fourier transform and its inverse transform respectively.  After the first phase screen, the random phase $\phi(x,y)$ is added to the electric field of the propagation beam in x-space, \begin{equation}
U_{1+}(x,y)=U_{1-}(x,y)exp(i\phi(x,y)),
\label{EQ_II}
\end{equation}
Such a procedure is continued until the last phase screens is reached.

The random phase screen $\phi(x,y)$ matching the fluctuations of the index of refraction could be described by a Fourier transform of a complex random number with variance $\sigma^2(k_x,k_y)$ , which can be expressed as,
\begin{equation}
\sigma^{2}(k_{x},k_{y})=\left(\frac{2\pi}{N\Delta x} \right)^{2}2\pi k_0^2 \Delta z \Phi(k_{x},k_{y}),
\label{EQ_III}
\end{equation}
where $N \times N$ is the size of phase screen, $k_0=2\pi/\lambda$ is the wave number of the light beam with wavelength $\lambda$, $\Delta x$ is the grid spacing in $x$ direction and the grid spacing in $x$ is assumed the same as that in $y$ direction, $\Delta Z$ is the spacing between the subsequent phase screens, and $\Phi(k_{x},k_{y})$ is the spectrum of the fluctuations in the index of refraction. Here, we use the form developed by Hill \cite{HI06} and defined analytically by Andrews \cite{AN98},
\begin{eqnarray}
\Phi(k_{x},k_{y}) & = & 0.033C_{n}^{2} [1+1.802\sqrt{\frac{k_{x}^2+k_{y}^2}{k_{l}^2}} \nonumber \\
& & -0.254[\frac{k_{x}^2+k_{y}^2}{k_{l}^2}]^{7/12}] \nonumber \\
& & \times \exp[{\frac{k_{x}^2+k_{y}^2}{k_{l}^2}}]
[k_{x}^2+k_{y}^2+\frac{1}{L_{0}^2}]^{-11/6},
\label{EQ_IV}
\end{eqnarray}
where $C_{n}^{2}$ is the structure constant of the index of refraction; $L_{o}^{2}$ is the outer scale of turbulence, that is the largest eddy size formed by injection of turbulent energy; $k_{l}=\frac {3.3}{l_{0}}$,  $l_{0}$ equals to the inner scale of turbulence; and  $k_i$ ($i =x,y$) is the wavenumber in $i$ direction. In fact, $C_{n}^{2}$ is a measure
of turbulence strength.
\subsection{Two correction methods in the proposed scheme}
Two correction methods are adopted in the proposed scheme. One is channel coding which is an optimal way to overcome noises in a communication link. BCH codes and Reed-Solomon (RS) codes are important error correction codes which have strong error correcting ability \cite{RE60}.  They are suitable for correcting multiple burst and random errors. Their cyclic properties make them easily shorten to any code length. For BCH codes, there are precise controls over the number of symbol errors correctable, and they can be decoded via an algebraic method, which simplifies the design of the decoder. Meanwhile, RS codes are a class of BCH codes, and are often viewed as cyclic BCH codes. RS codes are denoted by $(n, k, m)$, where $n$ is code length , $k$ is data length and $m$ is the bits for each RS symbol. The normal size of RS code is $n=2^m-1$.  

The other is wavefront correction method. The mitigation of `crosstalk' interference caused by AT could be obtained by including the retrieved phase $\varphi(x,y)$ which approximates  the distortion caused by AT to the propagation beam. Shack-Hartmann wavefront correction is a commonly used method to reconstruct random phase distortions caused by a aberration \cite{PL01}. Its principle is based on measurements of local slops of  Hartmann mask wavefront (Zernike polynomials), where Zernike polynomials are a set of orthogonal mathematical functions over a circle having a radius of $1$. In mathematics, any wavefront $\varphi(x,y)$ can completely  be described by a linear combination of Zernike polynomials $Z_0(x,y),Z_1(x,y),...,Z_N(x,y)$ and their tied coefficients $a_0,a_1,...,a_N$, that is,
\begin{equation}
\label{EQ_V}
\varphi(x,y)=\sum_{k=0}^\infty {a_k Z_k(x,y)}
\end{equation}
The coefficients ${a_k}$ could be obtained by calculating the partial derivation of the deformation wavefront and utilizing the least-squares method.

\section{Numerical simulations}
In this section, we will discuss the influence of AT on the OAM-multiplexed FSO communication links and the improvement of turbulence tolerance of the links by the proposed turbulence mitigation scheme. There exists, in principle, various optical beams that have OAM states. Here, Laguerre-Gaussian (LG) modes will be used and we will assume that there are four input bit streams at the sender.
\subsection{The turbulence mitigation effect by channel coding}
Firstly, we demonstrate the turbulence mitigation effect of our proposed scheme when channel coding is used. The turbulence aberration is simulated by $10$ uniformly spaced random phase screens. For each phase screen, the size $N \times N$ is setup to $128 \times 128$, the grid spacing $\Delta X $ and $\Delta Y$ are $0.002m$. The simulation parameters for the atmosphere turbulence are the following ones. The outer scale $L_0$  is  $50m$, the inner scale $l_o$ is  $0.001m$, the propagation distance is  $1000m$ and the space distance between each phase screen $\Delta Z$ is $100m$. The binary bits stream is represented by an independent and identically distributed (i.i.d.) binary sequence with $p_0 = p_1 = 1/2$.


\begin{figure}[!htbp]
\centering
\includegraphics{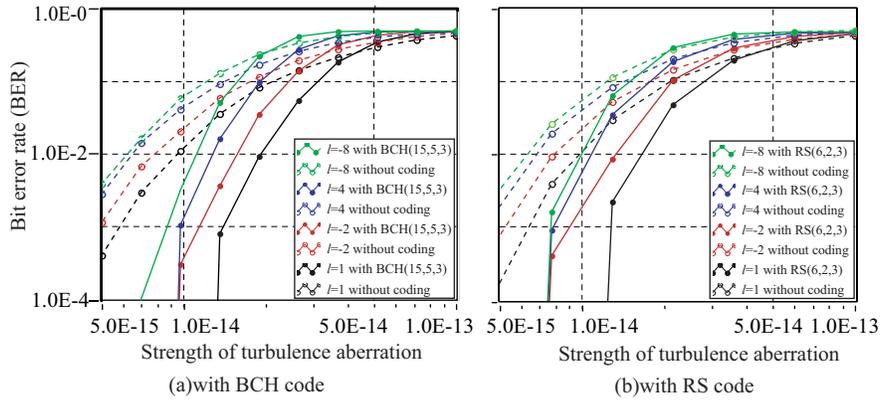}
\caption{BER performance of the OAM-multiplexed FSO communication
links under turbulence aberration with and without channel coding.}
\label{FIG_II}
\end{figure}
Fig.~\ref{FIG_II} shows the bit-error-rate (BER) performance of the OAM multiplexing FSO communication links with or without channel coding, where the four OAM states information-carriers are set to $\ell = \{+1,-2,+4,-8\}$.  
In order to illustrate the effect of the turbulent aberration strength  on the OAM-multiplexed FSO communication links, we give the results with the refractive index structure constant varying from $5.0 \times 10^{-15}m^{-2/3}$ to $10^{-13}m^{-2/3}$, representing the strength of AT from weak to strong \cite{AN08}. BCH(15,5,3) and RS(6,2,3) are individually  selected as the channel codes, and Berlekamp-Masssey (BM) algorithm is used to decode the codewords to obtain the output bits stream. Fig.~\ref{FIG_II}(a) shows the BER performance with or without BCH(15,5,3) channel codes, and Fig.~\ref{FIG_II}(b) presents the results with or without RS(6,2,3) channel codes.  The results show that the BER performance of the OAM-multiplexed FSO communication links decreases as the strength of turbulent aberration increases and the azimuthal mode index becomes larger. The BER performance has a big improvement when channel codes are used in the weak and medium turbulence. When the strength of atmospheric turbulence is less than $7.0 \times 10^{-15} m^{-2/3}$ ( weak turbulence), the errors in all azimuthal modes, including $\ell = \{+1,-2,+4,-8\}$, could be corrected by BCH(15,5,3) and RS(6,2,3).  However, the improvement of the performance by channel coding is limited by the correction ability of the selected channel codes. When the strength of atmospheric turbulence approaches to $5.0 \times 10^{-14}m^{-2/3}$ (relatively strong turbulence), there are no improvement in BER performance by using BCH(15,5,3)( see Fig.~\ref{FIG_II}(a))  and RS(6,2,3)( see Fig.~\ref{FIG_II}(b)), because the errors are beyond the correction ability of channel codes. When the strength of atmospheric turbulence reaches $1.0 \times 10^{-13}m^{-2/3}$ ( strong turbulence),  the BER performance for all azimuthal mode indexes $\ell$ approaches to $0.5$. This indicates that the interference caused by AT is very strong and the communication links has not ability to communicate.

On the other hand, the bigger azimuthal mode index $\ell$ OAM state is disturbed strongly by AT, and the improvement by channel coding is bigger for the smaller azimuthal mode index $\ell$ OAM state. When the strength of atmospheric turbulence is $9.0 \times 10^{-15}m^{-2/3}$, the azimuthal mode index ( $\ell=1$) OAM state has the smallest interference, the BER for $\ell=1$ OAM state is $1.0 \times 10^{-2}$ while it is $7.8 \times 10^{-2}$ for the azimuthal mode index $\ell=-8$ OAM state. 
The BER approaches infinity for $\ell=1$ OAM state with BCH(15,5,3) and RS(6,2,3), while it is $5.0 \times 10^{-3}$ with BCH(15,5,3) and $1.0 \times 10^{-2}$ with RS(6,2,3) for $\ell=-8$ OAM state, which has $15.6$-fold increase in the BER performance with BCH(15,5,3). Compared with RS(6,2,3) channel codes, BCH(15,5,3) codes have a better correction performance since they have the same correction ability and smaller code rate. 

\begin{figure}
\center
\includegraphics{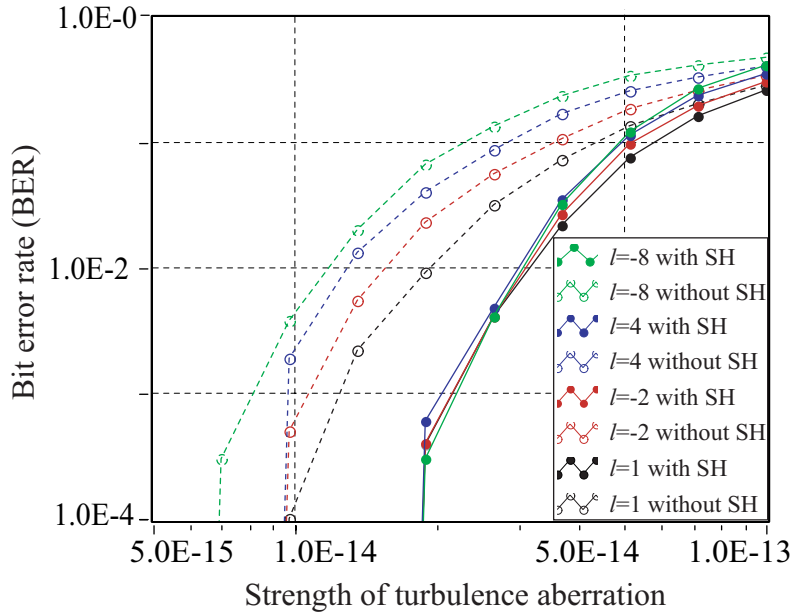}
\caption{BER performance of the OAM-multiplexed FSO communication
systems under turbulence aberration with and without Shack-Hartmann wavefront correction, where SH represents Shack-Hartmann wavefront correction method.}
\label{FIG_III}
\end{figure}
\subsection{The mitigation effect by a wavefront correction}
Secondly, we demonstrate the turbulence mitigation effect of our proposed scheme by a wavefront correction. The four OAM states carried bits streams are  set to $\ell = \{+1,-2,+4,-8\}$. The turbulence aberration is now simulated by one random phase screen, and the space distance is $500m$. The other parameters are the same as  above. It is shown that the probability to keep the original OAM state can be improved to $63.8\%$ by using Shark-Hartmann wavefront correction method \cite{ZH12}. In the OAM-multiplexed communication links, the information bits are modulated as the amplitudes of the four OAM states information-carriers. The higher fidelity of the OAM states at the receiver makes more accurate decisions on the carrying information, which results in the improvement of BER performance of the OAM-multiplexed communication links.

Fig.~\ref{FIG_III} shows the BER performance of the OAM-multiplexed FSO communication links with Shack-Hartmann wavefront correction, where $12^{th}$ order of Zernike polynomials are used to estimate the damaged wavefront by AT. The results show that the BER performances for all azimuthal OAM states  have been improved by using Shack-Hartmann wavefront correction.  The errors caused by AT will be corrected by using Shark-Hartmann wavefront correction  for all the OAM component states when the strength of atmospheric turbulence approaches to $7.0 \times 10^{-15}m^{-2/3}$, and the improvement by Shack-Hartmann wavefront correction is almost the same to the azimuthal mode indexes.

\subsection{An application of the turbulence mitigation scheme on images transmission}
In this subsection, we apply the proposed turbulence mitigation scheme to three gray-scale images' simultaneous transmission.
The three transmitted eight-level-gray-scale images are `Lena', `boat', and `NUPT' images. The `boat' image is carried by the azimuthal mode index $\ell=+1$ OAM state, the `Lena' image is carried by $\ell=0$ and the `NUPT' image is carried by $\ell=-1$. The phase screens representing AT is also obtained by simulation. The size $N \times N$ is setup to $128 \times 128$, the grid spacing $\Delta X $ and $\Delta Y$ are $0.002m$.  The outer scale $L_0$  is  $50m$, the inner scale $l_o$ is  $0.001m$, and the space distance $\Delta Z$ is $500m$. RS(6,2,3) is chosen for channel coding and Shack-Hartmann wavefront correction is used. 

In order to compare the quality of the reconstructed image quantitatively, the peak signal-to-noise (PSNR) is selected as an objective evaluation for the
imaging \cite{KO08}, which is defined as
\begin{equation}
\label{EQ_VI}
 MSE(a,b)  =\frac{1}{N\times M}{\sum_{x=0}^{N-1}}{\sum_{y=0}^{M-1}}(a(x,y)-b(x,y))^{2},
\end{equation}
and
\begin{equation}
 PSNR  =10log_{10}\left[\frac{maxVal^{2}}{MSE(a,b)}\right].
\label{EQ_VII}
\end{equation}
where $a(x,y)$ and $b(x,y)$ are the intensity values of the original
and the reconstructed image at  $(x,y)$, $N \times M$ is the size of the image, and $maxVal$ is the maximum possible pixel value of the image.
Generally, the higher PSNR is, the better quality the reconstructed
image has.
\begin{figure}
\centering
\includegraphics{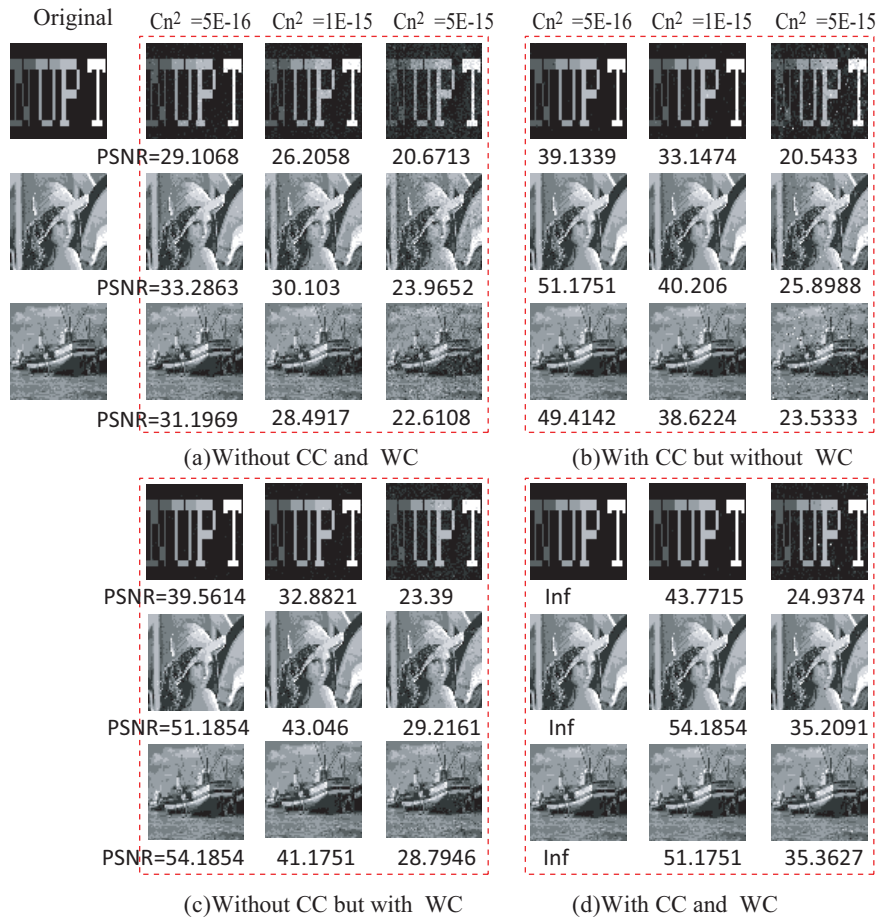}
\caption{Three eight-level-gray-scale images transmission over the OAM-multiplexed FSO communication links through AT, and the qualities are improved by the turbulence mitigation scheme. CC represents RS(6,2,3) channel code. WC means Shack-Hartmann wavefront correction method. }
\label{FIG_IV}
\end{figure}
\begin{figure}
\centering
\includegraphics{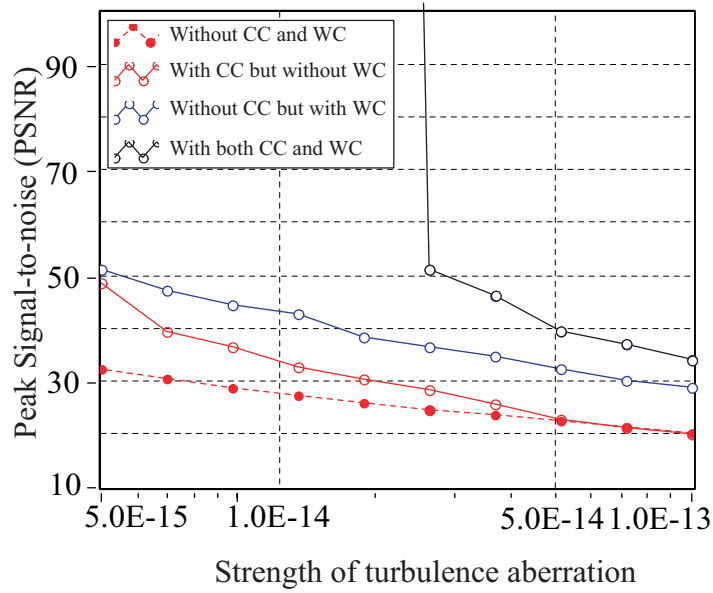}
\caption{The improvement to atmosphere turbulence tolerance by PSNR
against aberration strength. CC represents RS(6,2,3) channel code. WC means Shack-Hartmann wavefront correction method. }
\label{FIG_V}
\end{figure}
Fig.~\ref{FIG_IV} illustrates three eight-level-gray-scale images' transmission over the OAM-multiplexed FSO communication links through AT, and the quality improvements by the proposed turbulence mitigation scheme. In order to illustrate the effect of the strength of turbulent aberration on the OAM-multiplexed FSO communication links, the reconstructed images are presented with the refractive index structure constant  at $5.0 \times 10^{-16}m^{-2/3}$, $1.0 \times 10^{-15}m^{-2/3}$, and $5.0 \times 10^{-15}m^{-2/3}$, respectively. Meanwhile, the mitigation effect of the proposed scheme is displayed by the reconstructed images without channel coding and wavefront correction, with channel coding but without wavefront correction, with wavefront correction but without channel coding, and with channel coding and wavefront correction. Since the obtained phase screen is random even for the same simulation parameters, the PSNR is averaged over $10$ times in each cases. The results show that the qualities of the reconstructed images go down as the strength of turbulent aberration increases. Both channel coding and wavefront correction method could improve the qualities of the reconstructed images. The mitigation scheme that uses channel coding and wavefront correction method together can significantly enhance the qualities of the reconstructed images. For instance, the PSNR for `lena' is $33.2863$ at $C_n^{2}=5.0 \times 10^{-16}m^{-2/3}$, it could be completely recovered by the proposed mitigation scheme since the PSNR approaches infinity.

Fig.~\ref{FIG_V} shows how the PSNR values of the reconstructed
image with or without channel codes and wavefront correction
varies with the strength of atmosphere turbulence for the eight-level gray-scale image `Lena'. We let the index of refraction to change while keeping the other
simulation parameters to be constant to
calculate the PSNR for various strength of the turbulent aberration.
The PSNR in each cases is averaged over $20$
times. The results show that the
image quality could be improved by  channel coding and wavefront correction, and could be enhanced greatly by using channel coding and wavefront correction together. The PSNR approaches to infinity when $C_n^{2}$ is less than $2.8 \times 10^{-14}m^{-2/3}$. This indicates that the detrimental effect of AT for the OAM-multiplexed FSO communication links  could be removed by the proposed mitigation scheme when the strength of atmospheric turbulence even approaches to relatively strong turbulence.

\section{Conclusion}
In this paper, we have studied the method to improve the  atmospheric turbulence tolerance of the OAM-multiplexed FSO communication links using both channel coding and wavefront correction, where a set of random phase screens is adopted to simulate the turbulent aberration. In the scheme, 
we use channel codes to correct the burst and random errors on the individual OAM state introduced by AT. Simultaneously, we use Shark-Hartmann wavefront correction method to mitigate the phase distortion, which will produce `crosstalk' interference between the different OAM states. The improvements of AT tolerance of the OAM-multiplexed FSO communication links using channel coding and the wavefront correction method are analyzed and discussed by comparing the link with or without channel codes and wavefront correction method.  The results show that the BER performance of the OAM-multiplexed FSO communication links decreases as the strength of turbulent aberration increases, and the performances have improved greatly by using channel code and wavefront correction method.  When the proposed scheme is used in images transmission, the
image quality could be significantly improved  by using channel coding and wavefront correction together. The detrimental effect of AT on three images transmission could be removed by the proposed mitigation scheme when the strength of atmospheric turbulence even approaches to relatively strong turbulence.

%
\appendices
\section*{Acknowledgment}
We are grateful to Prof. Jozef Gruska for careful reading the manuscript. The work was supported in part by the National Natural Science Foundation of China (Grant No. 61271238 ), the Specialized Research Fund for the Doctoral Program of Higher Education of China (Grant No. 20123223110003), the University Natural Science Research Foundation of JiangSu Province (11KJA510002),  the open research fund of National Laboratory of Solid State Microstructure (M25020,M25022), and the project funded by the Priority Academic Program Development of Jiangsu Higher Education Institutions. L.Y. is partially supported by Qinglan project of Jiangsu Province.
\end{document}